\pdfoutput=1
\documentclass[
reprint,superscriptaddress,amsmath,amssymb,aps,prd,floatfix,footinbib]{revtex4-2}
\usepackage[a4paper,left=1.5cm,right=1.5cm,top=3cm,bottom=3cm]{geometry}
\usepackage[toc]{appendix} 
\usepackage{graphicx}
\usepackage{dcolumn}
\usepackage{bm}
\usepackage{amssymb,amsmath,amsfonts,physics,epsfig}
\usepackage[dvipsnames]{xcolor}
\usepackage{longtable}
\usepackage{verbatim}
\usepackage{color}
\usepackage{mdframed}
\usepackage{psfrag}
\usepackage{soul}
\usepackage{bm}
\usepackage{amsfonts,amssymb,mathrsfs,amsmath,esint}
\usepackage{slashed, cancel}
\usepackage{framed}
\usepackage{mdframed}
\usepackage{physics}
\usepackage{simplewick} 
\allowdisplaybreaks
\usepackage{latexsym}
\graphicspath{{./Figures/}}
\usepackage[dvipsnames]{xcolor}
\usepackage{booktabs}
\usepackage{datetime}
\newdateformat{mydate}{\THEDAY{ }\monthname[\THEMONTH]{ }\THEYEAR}

\usepackage{tikz}
\usepackage{color}
\usepackage{framed}
\usepackage{hyperref}
\usepackage[capitalise]{cleveref}
\hypersetup{colorlinks, citecolor=bluscuro, linkcolor=black, urlcolor=bluscuro}
\definecolor{rossos}{cmyk}{0,1,1,0.55}
\definecolor{bluscuro}{rgb}{0.15, 0.2, .85}
\definecolor{bluchiaro}{cmyk}{1,.3,0.,0.1}

\graphicspath{{./images/}}

\newcommand{\be}{\begin{equation}}
	\newcommand{\ee}{\end{equation}}
\newcommand{\bea}{\begin{eqnarray}}
	\newcommand{\eea}{\end{eqnarray}}
\newcommand{\beq}{\begin{equation}}
	\newcommand{\eeq}{\end{equation}}

\def\beqa{\begin{eqnarray}}

	\def\eeqa{\end{eqnarray}}

\def\lsim{\mathrel{\rlap{\lower4pt\hbox{\hskip0.5pt$\sim$}}
		\raise1pt\hbox{$<$}}}         
\def\gsim{\mathrel{\rlap{\lower4pt\hbox{\hskip0.5pt$\sim$}}
		\raise1pt\hbox{$>$}}}         

\makeatletter
\newcommand*{\rom}[1]{\expandafter\@slowromancap\romannumeral #1@}

\makeatother

\usepackage[normalem]{ulem}
\usepackage{soul}

\begin{document}
	\title{Spherically symmetric black holes in Gravity from Entropy and spontaneous emission}
	
	\author{Udaykrishna Thattarampilly}
	\email{uday7adat@gmail.com}
    \affiliation{Center for Gravitation and Cosmology, College of Physical Science and Technology, Yangzhou University, Yangzhou 225009, China}
   
	\author{Yunlong Zheng}
	\email{zhyunl@yzu.edu.cn (corresponding author)}

    	\affiliation{Center for Gravitation and Cosmology, College of Physical Science and Technology, Yangzhou University, Yangzhou 225009, China}
         \author{Vishnu Kakkat}
	\email{vishnu.kakkat@associated.ltu.se}
    \affiliation{Institutionen för teknikvetenskap och matematik, Luleå tekniska universitet, 971 87 Luleå, Sweden}
\affiliation{National Institute for Theoretical and Computational Sciences (NITheCS), South Africa}
	\date{\today}

	\begin{abstract}
		We investigate static and dynamical spherically symmetric black hole solutions within the Gravity from Entropy (GfE) framework. We derive and solve the modified vacuum field equations for a static, spherically symmetric spacetime, revealing that the classical Schwarzschild geometry receives perturbative corrections scaling as $r^{-4}$. We establish that the GfE framework is consistent with current strong-field astrophysical observations. 
          Higher-order geometric stresses inherent to the GfE vacuum drive a consistent mass-evolution profile. In the limit of large black hole mass, the theory predicts a constant background evaporation rate $ -\beta/24$, suggesting an inherent ``entropic leakage" of the vacuum. At intermediate scales, the framework replicates the standard Hawking radiation mass-loss law as $\dot{M} \propto M^{-2}$ through a purely classical response of the modified background.
  \end{abstract}
	
	\maketitle
\section{Introduction}	
The synthesis of general relativity and quantum information theory has emerged as an exciting field of contemporary theoretical physics, initiated by the seminal discovery that black holes possess an entropy proportional to the area of their event horizon \cite{Bekenstein:2008smd,Bekenstein:1974ax} and emit thermal radiation  \cite{Hawking:1975vcx}. These insights fundamentally imply that gravity inherently encodes quantum information, where entropy serves as a measure of the microscopic degrees of freedom inaccessible to a macroscopic observer. This information-theoretic perspective has been significantly deepened by the formulation of the holographic principle \cite{tHooft:1999rgb,Susskind:1994vu,Swingle:2009bg} and the subsequent advancements in  entanglement entropy \cite{Ryu:2006ef,Nishioka:2009un,Faulkner:2013ana,Witten:2018zxz,Sorce:2023fdx,Ben-Dayan:2023inz}. These developments challenge the classical view of gravity as a purely geometric construct, suggesting instead that gravitational dynamics may emerge as a thermodynamic or information-theoretic manifestation of underlying quantum degrees of freedom \cite{Padmanabhan:2009vy}.

A comprehensive information-theoretic approach to gravity is expected to provide deeper insights into the nature of the early universe, quantum gravity, and importantly for this work, black hole physics \cite{PhysRevLett.93.131301,Oriti_2009,Barack:2018yly}. Over the years, physicists have demonstrated that it is possible to derive the Einstein field equations from principles of thermodynamics and information theory \cite{Jacobson:1995ab,Verlinde:2010hp}. 
A promising study proposed in recent years has suggested that the quantum relative entropy between the spacetime metric and the metric induced by geometry and matter fields serves as the fundamental action governing the theory of gravity \cite{Bianconi:2024aju}. Quantum relative entropy is a central concept in information theory, defined for quantum operators in the context of von Neumann algebras \cite{Araki:1976zv,10.1093/oso/9780198517733.002.0001,ohya1993quantum,Vedral:2002zz}. In this recent approach dubbed Gravity from Entropy (GfE) framework, the spacetime metric, the geometry-induced metric, and the matter fields are treated as quantum operators, forming a bimetric theory of gravitation where metrics are promoted to renormalizable effective density matrices \cite{Rosen1973ABT,Hossenfelder:2008bg,Bianconi:2024aju}. This approach is shown to reduce to classical General Relativity in the weak-coupling and low-curvature limits and provides essential UV-complete descriptions at high energy scales \cite{Bianconi:2024aju}.

The GfE framework has recently been shown to yield naturally inflationary FLRW solutions without the need for additional scalar fields \cite{Thattarampilly:2025krv}. Implications of GfE theory for the strong-field regime of compact objects remain a vital area of inquiry and could open an avenue for observational tests. Recent work on approximate Schwarzschild solutions in entropic quantum gravity suggested that black holes with lengths far exceeding the Planck length obey the area law for entropy \cite{Bianconi:2025rnd}. However, a rigorous derivation of static, spherically symmetric solutions within the full modified vacuum equations of the GfE theory is required to understand the deviations from classical Schwarzschild geometry. 
In this article, we derive and solve the modified vacuum equations for a static, spherically symmetric spacetime. The entropic modifications effectively induce higher-derivative contributions to the gravitational dynamics, similar to the effective field theory approaches \cite{Donoghue:1994dn,Burgess:2003jk}. Specifically, our analysis reveals that the Schwarzschild geometry receives perturbative corrections scaling as $r^{-4}$, which lead to a deformation of the event horizon.

Crucially, we extend our analysis to the dynamical regime by solving the generalized field equations in comoving Lemaître coordinates. We find that the higher-order geometric stresses generated by the entropic modifications drive a consistent mass-evolution profile. Remarkably, for Black holes of mass higher than Plank mass, this approach replicates the standard Hawking radiation mass-loss law $\dot{M} \propto \frac{1}{M^2}$ 
 through a purely classical response of the modified background. Suggesting that spontaneous mass loss similar to Hawking radiation is intrinsic to spherically symmetric solutions of the theory. 
 
This paper is organized as follows: first, we discuss the Gravity from Entropy framework and the resulting modified vacuum equations. In the following sections, we solved these equations for a static, spherically symmetric ansatz and establish the asymptotic behavior of the metric. This is followed by a discussion of the observational constraints from S2 orbital data and the Event Horizon Telescope (EHT). Finally, we present the dynamic solution near the horizon and discuss the implications for black hole evaporation and spontaneous emission.
\section{Gravity from entropy Framework}
\label{sec:ge}
The fundamental premise of this work rests on the paradigm that gravitational dynamics are not merely an ontological property of spacetime geometry but emerge as a thermodynamic or information-theoretic consequence of quantum entanglement. Following the Gravity from Entropy approach proposed by Bianconi \cite{Bianconi:2024aju} and subsequently applied to cosmological inflation by Thattarampilly and Zheng \cite{Thattarampilly:2025krv}, we treat the spacetime metric $g_{\mu\nu}$ as a quantum operator. 
The gravity from entropy theory (Shortened as GfE theory) involves a topological metric composed of metrics between scalars, vectors, and bi-vectors defined on a 4-d manifold as 
\begin{equation}
    \tilde{g} = 1 \oplus g_{\mu \nu} dx^{\mu} \otimes dx^{\nu} \oplus g^{(2)}_{\mu \nu \rho \sigma} (dx^{\mu} \wedge dx^{\nu}) \otimes (dx^{\rho} \wedge dx^{\sigma}), 
\end{equation}
where $ g^{(2)}_{\mu \nu \rho \sigma} =\frac{1}{2} \left(g_{\mu \rho} g_{\nu\sigma}-g_{\mu\sigma} g_{\nu\rho} \right)$. To capture the interplay between the background geometry and matter fields, we introduce an additional induced metric $\tilde{\mathbf{G}}$ as 
\begin{equation}
    \tilde{\mathbf{G}} = \tilde{G}^{(0)} \oplus \tilde{G}^{(1)}_{\mu \nu} dx^{\mu} dx^{\nu} \oplus \tilde{G}^{(2)}_{\mu \nu \rho \sigma} (dx^{\mu} \wedge dx^{\nu}) \otimes (dx^{\rho} \wedge dx^{\sigma}), 
\end{equation}
where $\tilde{G}^{(0)}$, $\tilde{G}^{(1)}$ and $\tilde{G}^{(2)}$ are invertible at every point on the manifold. In this framework, $\tilde{g}$ and $\tilde{\mathbf{G}}$ act as renormalizable effective density matrices. 
The gravitational action is then identified with the \textit{Araki quantum relative entropy} between these two states.
The action proposed by GfE theory is given by 
\begin{equation}
    S = \frac{1}{\left(\textit{l}_{pl}\right)^4} \int \sqrt{-g} \mathcal{L} d^4 x 
    \label{eq:action}
\end{equation}
where $\textit{l}_{pl} = \left(\frac{\hbar G}{c^3} \right)^{1/2}$ is the plank length and $\mathcal{L}$, the Lagrangian density is 
\begin{equation}
    \mathcal{L} = -\mathrm{Tr} \log\left( \tilde{\mathbf{G}} \tilde{g}^{-1} \right).
    \label{eq:lagrangian}
\end{equation}
This logarithmic structure ensures that the theory reduces to classical General Relativity in the weak-coupling and low-curvature limits ($|GR| \ll 1$), while providing a UV-complete description at high energy scales. Variation of this action with respect to $g_{\mu\nu}$ leads to the modified Einstein equations.

In the vacuum regime, $\tilde{\mathbf{G}}$ is interpreted as a "dressed" metric, modified by the topological Ricci-Riemann tensor $\tilde{\mathcal{R}}$ \cite {Bianconi:2024aju,Bianconi:2025rnd}:
\begin{equation}
     \tilde{\mathbf{G}} = \tilde{g}-\frac{\beta }{2} \tilde{\mathbf{\mathcal{R}}} 
     \label{eq:gfie},
\end{equation}
where $\beta$ is the coupling parameter and 
\begin{equation}
     \tilde{\mathbf{\mathcal{R}}} = R \oplus (R_{\mu \nu}dx^{\mu }\otimes dx^{\nu} )\oplus R_{\mu\nu\rho\sigma} (dx^{\mu} \wedge dx^{\nu}) \otimes (dx^{\rho} \wedge dx^{\sigma}).
\end{equation}
$R$ is the Ricci scalar, $R_{\mu\nu}$ is the Ricci tensor, and $g^{\mu \eta} R_{\eta\nu\rho\sigma} $ is the Riemann tensor. To derive the black hole solutions presented in the following sections, we strictly adhere to the mathematical techniques established in the study of entropic inflation \cite{Thattarampilly:2025krv}. Specifically, in the absence of matter fields, we demanded that the Riemann curvature entries are non-zero only for non-summed diagonal indices. This allows us to map the bivector tensor $R^{(2)}_{\mu\nu\rho\sigma}$ onto a diagonal flattened $6 \times 6$ matrix, rendering the trace of the logarithmic action computationally tractable.
 By assuming a diagonal form for the Schwarzschild-like metric, we ensure that the Ricci tensor and the flattened Riemann matrix remain diagonal. This justifies the expansion of the Lagrangian in Equation as a sum of logs similar to the work done in \cite{Thattarampilly:2025krv}.

\section{Modified vacuum equations}  
For a static spherically symmetric spacetime in the absence of matter fields, the Ricci tensor and the Flattened Riemann tensor are diagonal. As in the inflationary case, we restrict our analysis to the special class of diagonal metrics. This assumption, as detailed in \cite{Thattarampilly:2025krv}, ensures that the flattened $6 \times6$
 matrix representation of the Riemann curvature remains diagonal, allowing the Lagrangian to be expressed as a sum of logarithmic terms:

\begin{equation}
\begin{split}
    -\mathrm{Tr} &\log\left( \mathbf{I}-\frac{\beta}{2} \tilde{\mathbf{\mathcal{R}}} \tilde{g}^{-1}   \right) = -\log\left(1- \frac{\beta}{2} R\right)-\\&\sum_{\mu\nu}^{\mu=\nu}\log\left( \delta_{\mu}^{\nu} -\frac{\beta}{2}R_\mu^{\nu}\right)-\sum^{\mu<\nu}_{\mu \nu} \log \left(
    \delta_{\mu\nu}^{\mu\nu}- \beta R_{\mu\nu}^{\;\;\;\mu \nu} \right) .
    \end{split}
      \label{eq:logr}
\end{equation}
Since both the flattened matrix and the Ricci tensor are diagonal, we can write
\begin{equation}
\begin{split}
   \mathbf{\mathcal{G}} &  = \frac{1}{(1-\frac{\beta}{2}R)} \oplus \frac{1}{\left( \delta_{\mu}^{\nu} -\frac{\beta}{2}R_\mu^{\nu} \right)}\; dx^{\mu} \otimes dx_{\nu} \\
    &\;\;\oplus \frac{1}{\left(\frac{1}{2}
    \delta_{\mu\nu}^{\rho\sigma} - \frac{\beta}{2} R_{\mu\nu}^{\;\;\;\rho \sigma}\right)}  (dx^{\mu} \wedge dx^{\nu}) \otimes (dx_{\rho} \wedge dx_{\sigma}) 
    \end{split}
\end{equation}
where $ \mathbf{\mathcal{G}}$ is defined as
\begin{equation}
     \mathbf{\mathcal{G}}^{-1} = \mathbf{I}-\frac{\beta}{2} \tilde{\mathbf{\mathcal{R}}} 
\end{equation}
The modified vacuum Einstein equation for the theory is given by \cite{Bianconi:2024aju}
\begin{equation}
    R^{\mathcal{G}}_{\mu\nu}-\frac{1}{2}  g_{\mu\nu}\left(R_{\mathcal{G}}-2\Lambda_{\mathcal{G}}\right)+D_{\mu\nu}=0
    \label{eq:modein}
\end{equation}
where
\begin{equation}
    R_{\mathcal{G}} =  -\mathrm{Tr} \left(g^{-1}\mathcal{G} \tilde{\mathcal{R}}\right),
\end{equation}
\begin{equation}
    \Lambda_{\mathcal{G}} =  -\frac{1}{2\beta}\mathrm{Tr} \left(\tilde{\mathcal{G}}- \tilde{\mathcal{I}}-\log\left(\tilde{\mathcal{G}}\right)\right),
\end{equation}
\begin{equation}
\begin{split}
R^{\mathcal{G}}_{\mu\nu} 
&= \mathcal{G}_{(0)}\, R_{\mu\nu} 
+ \left[ \mathcal{G}_{(1)} \right]^{\rho}{}_{\mu} R_{\rho\nu} \\
&\quad - \left[ \mathcal{G}_{(2)} \right]_{\rho_1 \rho_2 \mu \nu} R^{\rho_1 \rho_2} 
+ 2\, \left[ \mathcal{G}_{(2)} \right]^{\eta \rho_1 \rho_2}{}_{\mu} R_{\rho_1 \rho_2 \nu \eta}
\end{split}
\end{equation}
and
\begin{equation}
\begin{split}
D_{\mu \nu} &= \left( \nabla^{\rho} \nabla_{\rho} g_{\mu \nu}
- \nabla_{\mu} \nabla_{\nu} \right) \mathcal{G}_{(0)}
- \nabla^{\rho} \nabla_{\nu} \left[ \mathcal{G}_{(1)} \right]_{(\rho \mu)} \\
&\quad + \frac{1}{2} \nabla^{\rho} \nabla_{\rho} 
\left[ \mathcal{G}_{(1)} \right]_{\mu \nu}
+ \frac{1}{2} \nabla^{\eta} \nabla_{\eta} 
\left[ \mathcal{G}_{(1)} \right]_{\rho \nu} g_{\mu \nu} \\
&\quad + \nabla^{\eta} \nabla^{\nu} 
\left[ \mathcal{G}_{(2)} \right]_{\mu \rho \eta}
+ \nabla^{\eta} \nabla^{\nu} 
\left[ \mathcal{G}_{(2)} \right]_{\eta \mu \nu} \\
&\quad + \frac{1}{2} \left[ \nabla^{\rho}, \nabla^{\eta} \right] 
\left[ \mathcal{G}_{(2)} \right]_{\rho \eta \mu \nu}
\end{split}
\end{equation}
\section{Modified equations of motion} 
In this section, we solve the modified vacuum Einstein equations in a static, spherically symmetric setting. We adopt a static, spherically symmetric metric of the form
\[
ds^{2} = -A(r)\, dt^{2} + B(r)\, dr^{2} 
        + r^{2}\left(d\theta^{2} + \sin^{2}\theta\, d\varphi^{2}\right),
\]
where $A(r)$ and $B(r)$ are functions of the radial coordinate $r$ alone. 
Assuming the absence of matter fields, we substitute this line-element ansatz into equation \eqref{eq:modein}, thereby reducing the modified equations to three coupled ordinary differential equations.

The modified equations for entropic gravity are highly nonlinear and are given by
\begin{equation}
    \begin{split}
    R^{\mathcal{G}}_{00}+\frac{1}{2}  \left(R_{\mathcal{G}}-2\Lambda_{\mathcal{G}}\right)+D_{00}=0\\
     \frac{1}{g_{ii}} R^{\mathcal{G}}_{ii}+\frac{1}{2}  \left(R_{\mathcal{G}}-2\Lambda_{\mathcal{G}}\right)+\frac{D_{ii}}{g_{ii}}=0.
      \label{eq:modfried}
    \end{split}
\end{equation}
We refer the reader to the Appendix \ref{sec:app2}, for expanded expressions of all the terms in the equations up to first order in coupling $ \beta$. Equations concerning $ R^{\mathcal{G}}_{22}$ and $ R^{\mathcal{G}}_{33}$ are the same except for an overall multiplying factor. 
\\
\section{Corrections to schwarzchild metric in "Gravity from entropy"}
As we know, in the limit of standard General Relativity, these functions satisfy the Schwarzschild solution, i.e., $A(r) = B(r)^{-1} = 1 - 2M/r$ (in natural Units). However, the entropic corrections introduces perturbations to the field equations. The modified field equations, presented in detail in the Appendix (\ref{sec:app2}), contain terms proportional to the coupling parameter $\beta$. These terms represent the deviation from the classical Einstein-Hilbert action. A striking feature of these corrections is the appearance of higher-order derivatives of the metric functions, specifically up to the fourth derivative $B^{(4)}(r)$ and third derivative $A'''(r)$. This indicates that the entropic framework effectively induces non-local or higher-derivative contributions to the gravitational dynamics, characteristic of an effective field theory approach where the underlying microscopic degrees of freedom modify the macroscopic geometry at short length scales or high curvatures.

Moreover, physically, these corrections imply that the product $A(r)B(r)$ is no longer unity, violating the strong equivalence principle. 
The parameter $\beta$ controls the strength of the entropic modification.

\subsection{Asymptotic Series Expansion}

To determine the behavior of the metric at spatial infinity and to establish rigorous boundary conditions for numerical integration, we employ an asymptotic series expansion. We assume that far from the source ($r \to \infty$), the spacetime is asymptotically flat and approaches the Schwarzschild limit. Accordingly, we adopt the following form for the metric functions:
\begin{align}
    A(r) &= 1 - \frac{2M}{r} + \sum_{n=2}^{\infty} \frac{a_n}{r^n}, \\
    B(r) &= \frac{1}{1 - \frac{2M}{r}} + \sum_{n=2}^{\infty} \frac{b_n}{r^n},
    \label{eq:ser}
\end{align}
where $a_n$ and $b_n$ are coefficients to be determined by the modified field equations.

By substituting these expansions into the modified field equations (Eqs.~\ref{eq:eom4}--\ref{eq:eom1}) and solving order-by-order in powers of $1/r$, we determine the coefficients $\{a_n, b_n\}$.

We find that the lower-order coefficients vanish ($a_2 = a_3 = b_2 = b_3 = 0$), ensuring that the solution matches the Newtonian potential at large distances. The first non-trivial corrections due to the entropic coupling appear at order $\mathcal{O}(r^{-4})$. The explicit values for the leading coefficients are:
\begin{align}
    a_4 &= -\frac{ \beta M^2}{12}, & b_4 &= \frac{ \beta M^2}{3}, \\
    a_5 &= \frac{77  \beta M^3}{460}, & b_5 &= \frac{271  \beta M^3}{1380}.
\end{align}
This analytical series solution reveals that the entropic modifications induce a short-range deviation scaling as $M^2/r^4$ and $M^3/r^5$. These coefficients are used to define the initial values at a large radius $R_{\text{max}}$ for the numerical integration scheme, ensuring that the solution remains physical and asymptotically flat.

\begin{figure}[!ht]
    \centering
    \includegraphics[scale=0.65]{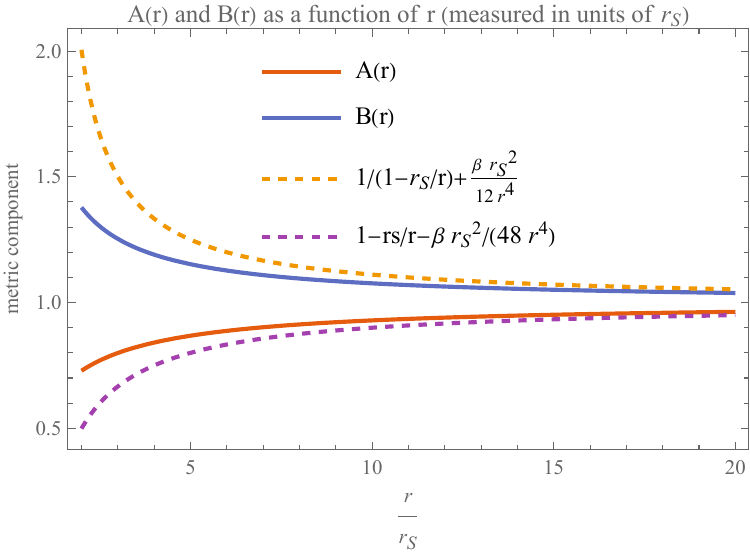}
    \caption{A(r) and B(r) as a function of $r$ (measured in units of Schwarzchild radius $r_S$) for $\beta =1$. The solid lines correspond to numerical evaluation of equations and the dashed lines represent the series solution in equation \eqref{eq:ser} truncated at the first correction to Schwarzchild solution.}
    \label{fig:metric}.
\end{figure}

The horizon can be located by setting
\[
A(r)=0 \implies 48 r^4-48 r_S r^3-\beta r_S^2=0,
\]
where $r_S$ is the Schwarzschild radius. The above equation admits a real, positive root, and the exact expression of which is given in the Appendix \ref{Appendix:static}. Expanding this root to first order in $\beta$ results in
\begin{equation}\label{eq:h_static}
 r_h=r_S+\frac{\beta}{48 r_S}+O(\beta^2).   
\end{equation}

Clearly, the location of the event horizon reduces to the Schwarzschild radius as $\beta\to 0$ as expected. Now, to determine whether the roots of the above equation correspond to physical horizons, we compute the Kretschmann scalar. The full expression for the Kretschmann scalar is provided in the Appendix \ref{Appendix:static}. From this, it is clear that except at $r=0$, all the singularities are removable, which leads to the conclusion that the surface located in eqn.\eqref{eq:h_static} is indeed represents an event horizon.

As seen in the series expansions utilized in our analysis, the corrections manifest primarily in the strong-field regime near the horizon radius, scaling with powers of the curvature. Consequently, the horizon structure is thermodynamically deformed, shifting the location of the event horizon and modifying the Hawking temperature relative to the standard black hole prediction.

The complexity of the system, particularly the mixing of non-linear terms with higher derivatives (e.g., terms of the form $A^3 (B')^4$), necessitates numerical integration to find global solutions matching asymptotic flatness. The solutions presented in Fig.~\ref{fig:metric} demonstrate that for small values of $\beta$, the metric functions smoothly interpolate between the modified near-horizon geometry and the Newtonian limit at infinity, preserving the causal structure while possibly introducing measurable corrections to gravitational lensing and perihelion precession at the scale of the coupling $\beta$.

\section{Deviation of Perihelion Shift from Schwarzschild}

To investigate the observational signatures of the modified gravity theory, we analyze the motion of test particles in the static, spherically symmetric spacetime defined by the metric:
\begin{equation}
    ds^2 = -A(r)dt^2 + B(r)dr^2 + r^2 d\Omega^2.
\end{equation}
The metric functions $A(r)$ and $B(r)$ contain a perturbative modification proportional to $r^{-4}$, parameterized by the dimensionless coupling constant $\beta$. In terms of the Schwarzschild radius $r_S = 2GM/c^2$, these are defined as:
\begin{equation}
    A(r) = 1 - \frac{r_S}{r} - \frac{\beta r_S^2}{48 r^4},
\end{equation}
\begin{equation}
    B(r) = \left(1 - \frac{r_S}{r}\right)^{-1} + \frac{\beta r_S^2}{12 r^4}.
\end{equation}
The standard Schwarzschild solution is recovered when $\beta \rightarrow 0$.

We derive the equation for the orbital precession (periapsis shift) by considering the geodesic equations for a particle with non-zero mass. Due to spherical symmetry, we confine the motion to the equatorial plane ($\theta = \pi/2$). The constants of motion, specific energy $\mathcal{E}$ and specific angular momentum $\mathcal{L}$, are governed by the radial equation:
\begin{equation}
    \left(\frac{dr}{d\tau}\right)^2 = \frac{1}{B(r)} \left[ \frac{\mathcal{E}^2}{A(r)} - \frac{\mathcal{L}^2}{r^2} - 1 \right].
\end{equation}
The trajectory of the orbit is bounded by two turning points, the periapsis $r_p$ and apoapsis $r_a$, where the radial velocity vanishes ($dr/d\tau = 0$). Solving the effective potential equation at these boundaries allows us to express $\mathcal{E}$ and $\mathcal{L}$ in terms of the metric functions evaluated at $r_p$ and $r_a$.

The orbital precession is determined by the total azimuthal angle $\Delta \phi$ accumulated during one complete radial period (from $r_p$ to $r_a$ and back). By converting the radial equation into an integral over $r$, we obtain:
\begin{equation}
    \Delta \phi = 2 \int_{r_p}^{r_a} \frac{d\phi}{dr} \, dr = 2 \int_{r_p}^{r_a} \frac{\mathcal{L}}{r^2} \sqrt{\frac{B(r)}{\frac{\mathcal{E}^2}{A(r)} - 1 - \frac{\mathcal{L}^2}{r^2}}} \, dr.
    \label{eq:int}
\end{equation}
The relativistic periapsis shift per orbit, $\delta \phi$, is the deviation from a closed Keplerian orbit:
\begin{equation}
    \delta \phi(\beta) = \Delta \phi - 2\pi.
\end{equation}
The integral in Eq.\eqref{eq:int}is evaluated numerically using standard quadrature methods to handle the coordinate singularities at the turning points.
\subsection{Near circular approximation}
We derived an analytical expression for the orbital precession using the near-circular (epicyclic) approximation. In this framework, the periapsis shift is determined by the discrepancy between the azimuthal frequency $\Omega_\phi$ and the radial epicyclic frequency $\Omega_r$ of a test particle on a stable circular orbit. 
By substituting the modified metric functions and expanding for small eccentricities and weak fields ($r_S/r \ll 1$), we map the circular radius $r_0$ to the semi-latus rectum $p = a(1-e^2)$ of the eccentric orbit. This expansion recovers the standard General Relativity prediction alongside the leading-order correction from the modified gravity term:
\begin{equation}
    \delta \phi \approx \frac{6\pi M}{c^2 a(1-e^2)} + \delta \phi_{\beta},
\end{equation}
where the first term is the standard Schwarzschild precession and $\delta \phi_{\beta}$ represents the additional shift contribution proportional to squire of the parameter $\beta$. 
\begin{equation}
    \delta \phi_{\beta} \approx \frac{\pi M \beta^2}{2 \left[ a(1-e^2) \right]^3}
\end{equation}
This analytical result is valid for orbits of small eccentricity. However, in order to observe a significant shift from Schwarzchild, we require observations of some highly eccentric orbits.

\subsection{Constraining $\beta$ using S2 Orbital Data}
We utilize the derived shift $\delta \phi$ to constrain the parameter $\beta$ against observational data from the star S2 (S0-2) orbiting Sagittarius A*. The GRAVITY Collaboration has quantified the relativistic precession via the Schwarzschild precession factor $f_{SP}$ \cite{GRAVITY:2020gka}, defined as:
\begin{equation}
    f_{SP} = \frac{\delta \phi(\beta)}{\delta \phi_{GR}},
\end{equation}
where $\delta \phi_{GR}$ represents the shift calculated in the Schwarzschild limit ($\beta=0$). The measured value is $f_{SP} = 1.10 \pm 0.19$.

\begin{figure}[!ht]
    \centering
    \includegraphics[scale=0.65]{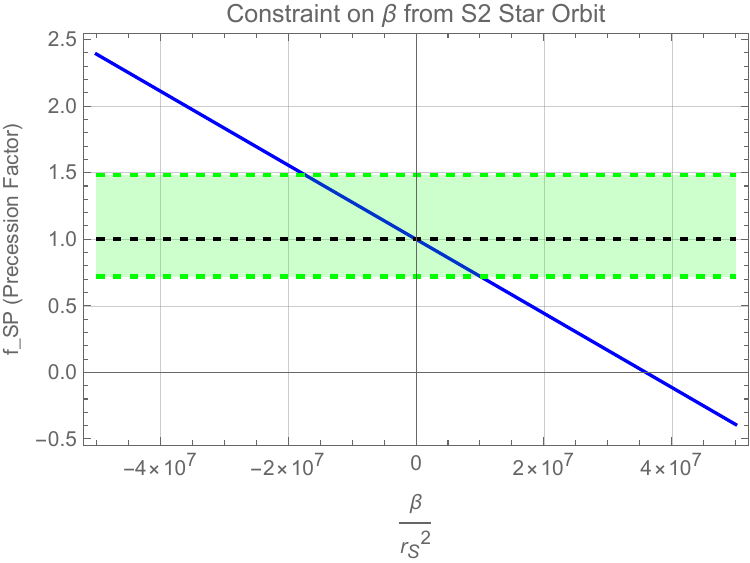}
    \caption{ $f_{SP} = \frac{\delta \phi(\beta)}{\delta \phi_{GR}}$ for the orbit of  S2 (S0-2) orbiting Sagittarius A* for different values of $\beta/r_S^2$. The shaded region is within the observed bounds. For all values of $-1<\beta<1$ theory is well within the observed bounds.}
    \label{fig:precession}.
\end{figure}

By computing $\delta \phi(\beta)$ for the specific orbital elements of S2 ($a \approx 12,500 r_S$, $e \approx 0.8846$), we determine the range of $\beta$ for which the theoretical prediction remains consistent with the $1\sigma$ and $2\sigma$ confidence intervals of the GRAVITY measurement. This comparison yields the observational bounds on the modification to the gravitational potential.
\section{Constraints from Black Hole Shadow}

While orbital precession tests the gravitational potential at distances of $r \sim 10^3 r_S$, the black hole shadow provides a direct probe of the spacetime geometry in the strong-field regime near the event horizon ($r \sim 1.5 r_S$). Given that the perturbative modification in our metric scales as $r^{-4}$, its magnitude is significantly enhanced at these small radii, making the shadow diameter a highly sensitive observable for constraining the parameter $\beta$.
\subsection{Shadow Diameter}
The boundary of the black hole shadow corresponds to the critical impact parameter $b_{crit}$ of photons that asymptotically approach the unstable photon sphere. For a static, spherically symmetric metric, the radius of the photon sphere, $r_{ps}$, is determined by the extrema of the effective potential for null geodesics. This yields the condition:
\begin{equation}
    \frac{d}{dr} \left( \frac{A(r)}{r^2} \right) \Bigg|_{r=r_{ps}} = 0 \quad \implies \quad r_{ps} A'(r_{ps}) - 2 A(r_{ps}) = 0.
\end{equation}
Substituting our modified metric function $A(r) = 1 - r_S/r - \beta r_S^2/(48 r^4)$ into Eq.~(9) results in an algebraic equation for the photon sphere radius. We solve this equation numerically for a given $\beta$ to find $r_{ps}$.

The critical impact parameter is related to the metric at the photon sphere by:
\begin{equation}
    b_{crit} = \frac{r_{ps}}{\sqrt{A(r_{ps})}}.
\end{equation}
For a distant observer at distance $D$ from the black hole, the angular diameter of the shadow, $\theta_{sh}$, is given in the small-angle approximation by:
\begin{equation}
    \theta_{sh} = 2 \frac{b_{crit}}{D}.
\end{equation}
In the limit $\beta \to 0$, this recovers the standard General Relativity prediction of $\theta_{sh} \approx 6\sqrt{3} GM / (c^2 D)$.

\subsection{Observational Bounds from EHT}
We compare our theoretical predictions against the shadow diameter of Sagittarius A* measured by the Event Horizon Telescope (EHT) Collaboration. We utilize the angular diameter $\theta_{obs} = 51.8 \pm 2.3~\mu\text{as}$ \cite{EventHorizonTelescope:2022wkp}, adopting the distance $D \approx 8178~\text{pc}$ and mass $M \approx 4.154 \times 10^6 M_\odot$.
\begin{figure}[!ht]
    \centering
    \includegraphics[scale=0.42]{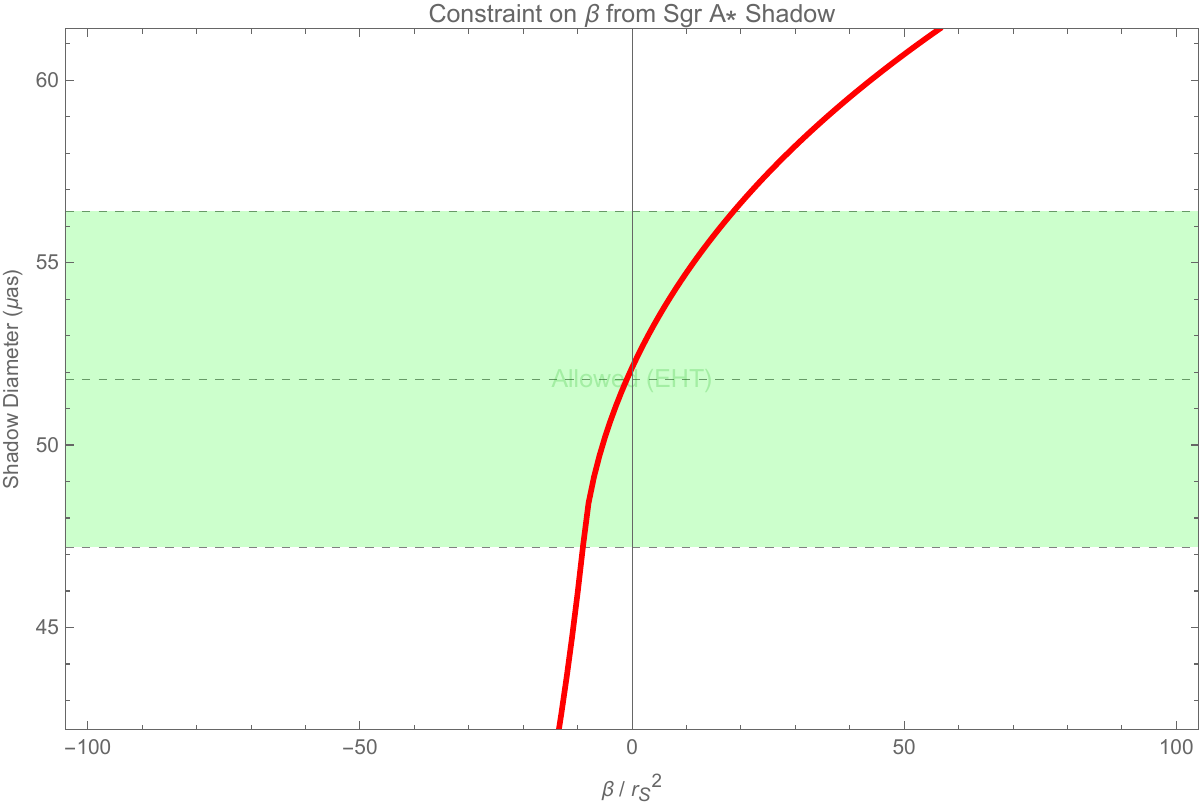}
    \caption{Prediction for shadow diameter of Sagittarius A* as a function of parameter $\beta$. The shaded region represent observed shadow diameter measured by the Event Horizon Telescope (EHT) Collaboration. }
    \label{fig:shadow}.
\end{figure}

We computed $\theta_{sh}(\beta)$ for a range of dimensionless $\beta$ values. Unlike the orbital precession, which required $\frac{\beta}{r_S^2} \sim 10^{5}$ to manifest observable deviations, the shadow diameter shows significant sensitivity to $\beta$ of order unity. By requiring the calculated shadow size to remain within the $2\sigma$ confidence interval of the EHT measurement ($47.2~\mu\text{as} \leq \theta_{sh} \leq 56.4~\mu\text{as}$), we derive the following constraint on the modified gravity parameter:
\begin{equation}
    -9.04 r_S^2\leq \beta \leq 18.63 r_S^2.
\end{equation}
This tight constraint confirms that strong-field lensing observables provide a superior test for gravitational modifications scaling with high inverse powers of distance.
 \section{Dynamic solution near the horizon and Black hole evaporation}
The static modified metric, while useful for establishing asymptotic corrections far from the source, becomes mathematically insufficient as one approaches the event horizon. To evaluate the behavior of solutions near Horizon, we must deviate from our static ansatz. 
To allow for the dynamical evolution of black holes within our GfE framework, we solve the generalized Einstein field equations, $G_{\mu\nu} = -\beta  H_{\mu\nu}$, where  $H_{\mu\nu}$ encapsulates the higher-order geometric corrections. Given the requirement for a strictly diagonal Ricci tensor in a dynamic setting, we adopt the generalized Lemaître (comoving) metric ansatz:
\begin{equation}
    ds^2 = -g(t, r)^2 dt^2 + \left( \frac{\partial f(t, r)}{\partial r} \right)^2 dr^2 + f(t, r)^2 d\Omega^2,
    \label{eq:metric2}
\end{equation}
where $g(t, \rho)$ is the lapse function (gravitational redshift) and $f(t, r)$ is the areal radius. A similar approach is established in \cite{Bianconi:2025awa}, where the author discusses how thermodynamic quantities for FLRW Universe are related to the microscopic degrees of freedom of geometry.
\subsection{Derivation of the Equations of Motion}
The derivation proceeds by first formulating the modified gravity source terms $H_{\mu\nu}$ under the assumption of a vanishing lapse perturbation ($g=1$). The field equations derived this way are described in appendix \ref{sec:app3}. While calculating field equations, no specific constraints are placed on the areal radius $f(t, \rho)$. We calculate the modification terms assuming that the spacetime is nearly Schwarzschild. For Schwarzchild spacetime $f_{Sch}(t, r) = \left[ \frac{3}{2} \sqrt{2M} (r - t) \right]^{2/3}$. where $M$ is the mass of the black hole. Substituting this in modified equations \eqref{eq:mo1},\eqref{eq:mo2},\eqref{eq:mo3} and
 evaluating the three modification components at the apparent horizon ($f = 2M$) yields the following source terms:
\begin{align}
    \mathcal{H}_{tt} &= -\beta  \left( \frac{1 + 4 M (7 + 18 M)}{256 M^5}\right), \label{eq:Htt} \\
    \mathcal{H}_{\rho\rho} &= \beta \left( - \frac{11}{4} + \frac{1722 - 2071 M}{13440 M^3}\right), \label{eq:Hrr} \\
    \mathcal{H}_{\theta\theta} &= \beta  \left( \frac{-1265 + 46029 M^2 + 11040 M^4}{176640 M^6} \right). \label{eq:Htheta}
\end{align}
These expressions represent the fixed geometric "stresses" generated by the higher-order curvature modifications of the theory. While these terms are derived from a static background, they act as the driving force for the dynamical response of the spacetime.

Subsequently, we introduce the dynamical degrees of freedom into the Einstein tensor $G_{\mu\nu}$. Specifically, we allow the mass parameter to become time-dependent, following the perturbative ansatz:
\begin{equation}
    M(t) = M_0 + \beta  m(t),
\end{equation}
where $M_0$ is the static Schwarzschild mass and $m(t)$ is the first-order dynamical correction. Simultaneously, we re-introduce the lapse function $g(t, \rho)$ into the Einstein part of the field equations \ref{sec:app3}. While $g=1$ was used to define the source $H_{\mu\nu}$, the Einstein tensor must account for $g \neq 1$ to provide the mathematical flexibility required to satisfy the transverse pressure and energy density constraints simultaneously.  This approach is justified by the order of the coupling; since $H_{\mu\nu}$ is already linear in $\beta $, the inclusion of first-order dynamical perturbations within these terms would contribute only at $\mathcal{O}(\beta ^2)$, which is neglected in this linear-order analysis. Einstein equations $G_{\mu\nu}$ in Lemaitre coordinates upon introducing the lapse function is given in appendix \ref{sec:app31}. 
We solve the field equations assuming that the mass and lapse functions are varying quasi-statically.  
 By equating these stationary source terms to the Einstein tensor $G_{\mu\nu}$—which explicitly includes the first-order time-dependent mass $M(t)$ and the lapse perturbation $h(t, \rho)$—we resolve the system of equations to find the consistent mass-evolution law. 
\subsection{The Perturbed Field Equations}
In the Lemaître coordinate system, the linearized Einstein tensor components at the apparent horizon take the following form (omitting higher-order adiabatic terms):
\begin{align}
    \delta G_{tt} &\approx \beta  \frac{3 \dot{m}}{2 M^2}, \label{eq:pertGtt} \\
    \delta G_{\theta\theta} &\approx \frac{\beta ( \dot{m} + 2 h'(t,r)-2M h''(t,r))}{ M^2}\\
    \delta G_{rr} &\approx  \frac{\beta ( \dot{m} +  h'(t,r))}{ M^2}
    \label{eq:pertGtheta}
\end{align}
where $\dot{m} \equiv \beta  \dot{M}$ represents the mass-loss rate and $h'$ denotes the derivative of h w.r to co $\tau=r-t$. By equating these linearized geometric terms to the previously derived source terms $\mathcal{H}_{\mu\nu}$, we establish a consistent system of equations for the dynamical variables.

\subsection{Coupling of Mass Change and Gravitational Redshift}

The simultaneous solution of the perturbed equations reveals a relation between the evaporation rate and the gravitational lapse. From the off-diagonal and transverse components, we find that the consistency of the system requires:
\begin{equation}
    h'(t, r) \sim - \dot{m}. \;\;\;\;\;\;\; h''(t,r) \sim -\frac{1}{2M}
\end{equation}
This result demonstrates that the lapse function is not a passive coordinate choice but an active participant in the mass-evolution process. Physically, the lapse gradient represents the adjustment of the gravitational redshift required to maintain energy conservation as the black hole loses mass. Without this non-trivial evolution of the temporal flow, the radial stresses and energy density of the modified background would remain inconsistent, effectively prohibiting any mass change under Birkhoff's theorem.

The resulting  equation for the mass evolution, obtained by evaluating the temporal field equation \eqref{eq:Htt} against the perturbed Einstein component, is:
\begin{equation}
    \dot{M} = \beta  \left(-\frac{1}{24}-\frac{0.17}{M^2}+\frac{0.005}{M^4} \right).
    \label{eq:ma}
\end{equation}
This equation signifies that the mass loss is driven by the imbalance between the background vacuum curvature (the constant terms) and the localized horizon curvature (the $M^{-2}$ terms). The perturbative approach confirms that while the source of the modification is calculated from a static background, the spacetime responds by shifting its temporal and mass scales in a synchronized manner, leading to the geometric evaporation profile observed.
\subsection{ Mass Loss}
To analyze the phenomenological behavior of this result, we perform an asymptotic expansion for the limit of large black hole mass ($M \gg 1$):
\begin{equation}
    \dot{M} \approx -\frac{\beta }{24} - \beta \frac{0.17}{ M^2} + \mathcal{O}\left(\frac{1}{M^4}\right).
\end{equation}
The derived mass-loss law reveals a dual-regime evaporation profile. The constant term, $\dot{M} \to -\beta /24$, suggests a background geometric "decay" inherent to the modified vacuum. More significantly, the $M^{-2}$ term provides a, geometric replication of the Hawking radiation profile. 

This term signifies a background geometric decay inherent to the modified vacuum of the theory. Unlike standard Hawking radiation, which vanishes as mass increases, this constant loss rate implies that the black hole is "leaking" energy into the entropic fabric of spacetime at a steady rate, regardless of its size. In an information-theoretic context, this represents the dissipation of the "G-field" stresses that exist even in the limit of low curvature ($M\gg1$), suggesting that the vacuum itself in GfE is not strictly stationary but possesses an underlying dissipative flow. This background decay is similar to the mass decrease observed in black holes accreting phantom energy \cite{Babichev:2004yx,Gao:2008jv} or those embedded in dynamical de Sitter backgrounds \cite{Sultana:2005tp,Giulini:2013zha}. This is not surprising, since the cosmological constant is inherent to GfE theory. 
While a constant mass-loss term is absent in classical General Relativity, it also appears in several modified/quantum gravity frameworks \cite{Gregory:2017ogk,Barrow:1997qh,Cisterna:2014nua}. 
\subsection{Thermodynamics of Spontaneous Emission from Full Mass Loss}
The total power radiated by the black hole is given by 
\begin{equation}
    P = -\dot{M} = \beta \left( \frac{1}{24} + \frac{0.17}{M^2} - \frac{0.005}{M^4} \right)
\end{equation}
To define the thermodynamic variables, we assume this power is emitted thermally across the horizon area $A = 4\pi r_h^2$. For a Schwarzschild-like object in this theory, the leading-order area is $A \approx 16\pi M^2$.
Equating the GfE power loss to the Stefan-Boltzmann law, $P = \sigma A T^4$, we solve for the effective temperature $T$:
\begin{equation}
    T(M) = \left( \frac{\beta}{16\pi\sigma} \right)^{1/4} \left( \frac{1}{24 M^2} + \frac{0.17}{M^4} - \frac{0.005}{M^6} \right)^{1/4}.
\end{equation}
In the limit of large black hole mass ($M \gg 1$), we expand the temperature expression to obtain:
\begin{equation}
    T(M) \approx \left( \frac{\beta}{384\pi\sigma} \right)^{1/4} \frac{1}{\sqrt{M}}.
\end{equation}
While the temperature vanishes as $M \to \infty$, it follows a $T \propto M^{-1/2}$ scaling law. This is a significant departure from the standard Schwarzschild Hawking temperature, $T_H \propto M^{-1}$. This implies that macroscopic black holes in GfE theory remain warmer than those in General Relativity, as the ``entropic leakage" prevents the temperature from dropping as rapidly as mass increases.
\begin{equation}
    C(M) \simeq \frac{64\pi\sigma T^3}{\beta} \left( -\frac{1}{12 M^3} - \frac{0.68}{M^5} + \frac{0.03}{M^7} \right)^{-1}
\end{equation}
For large $M$, the leading order term is:
\begin{equation}
    C(M) \approx -\frac{768\pi\sigma T^3 M^3}{\beta}
\end{equation}
The heat capacity remains negative, confirming that GfE black holes are thermodynamically unstable. However, the $M^{-5}$ and $M^{-7}$ corrections imply that the rate of instability is modified at small scales, potentially altering the final stages of the evaporation process.

 \section{Conclusions and discussions}
   We have investigated the properties of static and dynamical spherically symmetric spacetimes within the Gravity from Entropy (GfE) framework. Our analysis of the static solution reveals that the Schwarzschild geometry receives perturbative corrections scaling as $r^{-4}$, which lead to a physical deformation of the event horizon. Specifically, the horizon radius is shifted from its classical value to $r_h \approx r_s + \beta/(48r_s)$, where $\beta$ represents the entropic coupling parameter. This shift indicates that the microscopic degrees of freedom encoded in the gravitational action effectively modify the macroscopic geometry at high curvatures, a hallmark of effective field theory approaches to quantum gravity.

We further tested these theoretical predictions against observational data from the S2 star's orbital precession and the Event Horizon Telescope's shadow measurements of Sagittarius A*. While the periapsis shift of the S2 star provides a broad constraint on the modification to the gravitational potential, the shadow diameter serves as a far more sensitive probe of the near-horizon geometry. By requiring the calculated shadow size to remain within the $2\sigma$ confidence interval of the EHT measurement, we established the constraint $-9.04r_s^2 \leq \beta \leq 18.63r_s^2$. This result demonstrates that the GfE framework is consistent with current strong-field astrophysical observations.

More interestingly, our equations predict the emergence of a spontaneous mass-loss law from a purely classical treatment of the modified background in Lemaître coordinates. We found that the higher-order geometric stresses drive a mass-evolution profile $\dot{M} \approx -\frac{\beta}{24} - \beta \frac{0.17}{M^2}$, which replicates the standard Hawking radiation mass-loss law for black holes of mass higher than the Planck mass. This ``geometric evaporation" suggests that Hawking-like radiation is an intrinsic property of the modified vacuum in GfE theory, rather than requiring a separate treatment of quantum fields on a fixed background. Furthermore, the resulting temperature scaling $T \propto M^{-1/2}$ represents a significant departure from the standard $T \propto M^{-1}$ relation, implying that macroscopic black holes in this framework remain warmer than their classical counterparts, potentially impacting our understanding of black hole lifetimes and the final stages of evaporation. 

The modified mass-evolution law derived in Eq.\eqref{eq:ma} suggests a significant modification to the final stages of the evaporation process. If the equations leads to a vanishing mass-loss rate at a finite, non-zero mass, the GfE framework would naturally predict the existence of stable black hole remnants. Such a result would have profound implications for the resolution of the black hole information paradox, as these remnants could potentially serve as long-lived repositories for the quantum information encoded in the original matter distribution. Establishing the existence of such remnants would position the GfE framework as a robust candidate for a UV-complete theory of gravity that avoids the formation of naked singularities.
\\

    \textit{\textbf{Acknowledgments}}-- We would like to acknowledge Prof. Ginestra Bianconi (Queen Mary University of London) for her insights and suggestions that helped to improve the draft significantly. This work is supported in part by NSFC under Grant No. 11847239. VK is supported by the Kempe Foundation grant JCSMK24-507.

    \newpage
	
	\bibliography{ref.bib}
	
	\newpage
	\pagebreak

    \newpage
    \widetext
\begin{center}
\end{center}

\appendix
    \section{Modified Einstein equations for static spherically symmetric space time}   \label{sec:app2}
    By substituting the metric anzats in equation \eqref{eq:modein} we obtain the modified equations of motion\eqref{eq:modfried}. Assuming that the coupling $\beta$ is small we expand the equations in $\beta $ and truncate them at first order. The equations of motion \eqref{eq:modfried} expanded up to first order in $\beta $ are given by

\begin{equation}
\begin{split}
\mathrm{eom}_{tt} & = \frac{3 B (A^2 + r A' - A)}{A^2 r^2} + \frac{ \beta}{128 A^5 B^3 r^4} \Bigg[ 
96 A^3 B^{(4)} B^3 r^4 - 288 B^2 A^2 r^3 \left(r A B' + B \left(r A' - \frac{4 A}{3}\right)\right) B''' \\
& \quad - 48 B^3 A^2 r^3 \left(r B' + \frac{2 B}{3}\right) A''' - 212 (B'')^2 B^2 A^3 r^4 \\
& \quad - 192 B A r^2 \left( A B^2 A'' r^2 - \frac{173 (B')^2 A^2 r^2}{48} - \frac{161 B A \left(r A' - \frac{30 A}{23}\right) r B'}{48} - \frac{19 B^2 \left((A')^2 r^2 - \frac{91 r A' A}{57} - \frac{4 A^2 (A - 1)}{57}\right)}{8} \right) B'' \\
& \quad + 312 B^2 A \left(\frac{6 (B')^2 A r^2}{13} + B r \left(r A' - \frac{2 A}{3}\right) B' + \frac{2 B^2 \left(r A' - \frac{2 A}{13}\right)}{3}\right) r^2 A'' - 293 A^3 (B')^4 r^4 \\
& \quad - 346 B A^2 r^3 \left(r A' - \frac{218 A}{173}\right) (B')^3 - 341 B^2 A r^2 \left((A')^2 r^2 - \frac{512 r A' A}{341} - \frac{16 A^2 (A + 1/4)}{341}\right) (B')^2 \\
& \quad - 336 B^3 r \left(r^3 (A')^3 - \frac{41 r^2 A (A')^2}{28} - \frac{A^2 (A + 11/2) r A'}{21} + \frac{5 A^3 (A - 1)}{21}\right) B' \\
& \quad - 224 B^4 \left(r^3 (A')^3 - \frac{61 r^2 A (A')^2}{56} - \frac{13 A^2 (A - 9/13) r A'}{14} - \frac{15 A^3 (A - 1) (A - 19/15)}{14}\right) \Bigg] =0
\end{split}
\label{eq:eom4}
\end{equation}

\begin{equation}
\begin{split}
\mathrm{eom}_{\theta \theta} & = -\frac{3 r \left(B' A' r B - 2 B'' A r B + (B')^2 A r + 2 A' B^2 - 2 B' A B\right)}{4 A^2 B^2} \\
& \quad + \frac{G \beta}{128 A^5 B^4 r^2} \Bigg[ -16 A^3 B^3 B^{(4)} r^4 + \left(40 r^4 B^2 A^3 B' + 48 B^3 r^3 \left(A' r - \frac{5 A}{3}\right) A^2\right) B''' \\
& \quad + 8 r^3 B^3 A^2 (B' r + 14 B) A''' - 28 (B'')^2 A^3 B^2 r^4 + 32 B r^2 \Bigg( (A'' A B^2 r^2) - \frac{9 (B')^2 A^2 r^2}{8} \\
& \quad - B \left(A' r - \frac{11 A}{2}\right) r A B' - \frac{19 B^2 \left((A')^2 r^2 - \frac{68 A' r A}{19} - \frac{8 A^2 (A - 2)}{19}\right)}{8} \Bigg) A B'' \\
& \quad - 52 B^2 \left(\frac{5 (B')^2 A r^2}{13} + r B \left(A' r - \frac{38 A}{13}\right) B' + 14 B^2 \left(A' r - \frac{8 A}{91}\right)\right) r^2 A A'' \\
& \quad + 25 A^3 (B')^4 r^4 + 18 B \left(A' r - \frac{58 A}{9}\right) r^3 A^2 (B')^3 \\
& \quad + 33 B^2 \left((A')^2 r^2 - \frac{64 A' r A}{11} - \frac{16 A^2 (A - 1/4)}{33}\right) r^2 A (B')^2 \\
& \quad + 56 B^3 \left(r^3 (A')^3 - \frac{95 A r^2 (A')^2}{14} - \frac{2 (A - 13/2) r A^2 A'}{7} - 2 A^4 + \frac{18 A^3}{7}\right) r B' \\
& \quad + 784 B^4 \left(r^3 (A')^3 - \frac{47 A r^2 (A')^2}{196} + \frac{A^2 r (A - 3) A'}{7} + \frac{5 A^3 (A + 3) (A - 1)}{49}\right) \Bigg] =0
\end{split}
\label{eq:eom2}
\end{equation}

\begin{equation}
\begin{split}
\mathrm{eom}_{rr} & = \frac{3 B' r - 3 B (A - 1)}{r^2 B} + \frac{ \beta}{128 A^4 B^4 r^4} \Bigg[ -16 (3 B' A r + B (r A' + 2 A)) B^2 r^3 A^2 B''' \\
& \quad + 20 (B'')^2 A^3 B^2 r^4 + 24 B \Bigg(\frac{19 (B')^2 A^2 r^2}{6} + \frac{7 B r \left(r A' - \frac{6 A}{7}\right) A B'}{2} \\
& \quad + B^2 \left((A')^2 r^2 - r A' A + \frac{4 A^3}{3} - 4 A^2\right)\Bigg) r^2 A B'' + 8 \left(3 (B')^2 A r^2 + r B (r A' + 4 A) B' + 4 B^2 (r A' + 7 A)\right) B^2 r^2 A A'' \\
& \quad - 43 A^3 (B')^4 r^4 - 54 \left(r A' - \frac{22 A}{27}\right) B r^3 A^2 (B')^3 \\
& \quad - 59 B^2 r^2 \left((A')^2 r^2 - \frac{96 r A' A}{59} + \frac{16 (A - 19/4) A^2}{59}\right) A (B')^2 \\
& \quad - 16 \left(r^3 (A')^3 + \frac{3 r^2 A (A')^2}{4} + A^2 r \left(A - \frac{33}{2}\right) A' - 13 A^4 + 9 A^3\right) B^3 r B' \\
& \quad - 64 B^4 \left(r^3 (A')^3 + \frac{97 r^2 A (A')^2}{16} + \frac{9 A^2 r (A - 1) A'}{4} + \frac{15 A^5}{4} - \frac{A^4}{2} - \frac{13 A^3}{4}\right) \Bigg] =0
\end{split}
\label{eq:eom1}
\end{equation}

\section{Horizon and the Kretschmann scalar for the static metric}
\label{Appendix:static}
To express the static horizon radius compactly, let

\[
\xi
\equiv
\sqrt[3]{
\sqrt{\beta^2 r_S^{6}\left(16\beta+81 r_S^{2}\right)}
-9\beta\,r_S^{4}
}.
\]

The horizon radius can then be written as
\begin{equation}\label{eq:horizon}
\begin{aligned}
r_h &= \frac{1}{12} \Bigg[
3\,r_S \;+\;
\sqrt{3}\, 
\sqrt{
2^{1/3}\,\xi
+
r_S^2 \Big(3 - \frac{2\,2^{2/3}\beta}{\xi}\Big)
} \;+\;
\\[2mm]
&\qquad
\sqrt{3}\,
\sqrt{
6 r_S^2
- 2^{1/3}\,\xi
+ \frac{2\,2^{2/3}\beta\,r_S^2}{\xi}
+ \frac{6\sqrt{3}\,r_S^3}{
\sqrt{2^{1/3}\,\xi
+
r_S^2 \Big(3 - \frac{2\,2^{2/3}\beta}{\xi}\Big)}
}
}
\Bigg].
\end{aligned}
\end{equation}

Clearly, the above equation reduces the Schwarzschild radius in the limit
\(\beta\to0\).

We can calculate the Kretschmann scalar for the static metric as
\begin{align}
K &= \frac{4 r_S^2}{r^4 \left(\beta r_S^2 + 48 r^3 (r_S-r)\right)^4 \left(\beta r_S^2 (r_S-r) + 12 r^5\right)^4} \times \Biggl(
\beta^8 r_S^{14} (r-r_S)^4 \notag \\
&\quad + 24 \beta^7 r^3 r_S^{12} (r_S-r)^3 \bigl(9 r^2 - 15 r r_S + 8 r_S^2\bigr) \notag \\
&\quad + 144 \beta^6 r^6 r_S^{10} (r-r_S)^2 \bigl(149 r^4 - 452 r^3 r_S + 565 r^2 r_S^2 - 352 r r_S^3 + 96 r_S^4\bigr) \notag \\
&\quad + 3456 \beta^5 r^9 r_S^8 (r_S-r) \bigl(396 r^6 - 1647 r^5 r_S + 2925 r^4 r_S^2 - 2904 r^3 r_S^3 + 1776 r^2 r_S^4 - 672 r r_S^5 + 128 r_S^6\bigr) \notag \\
&\quad + 10368 \beta^4 r^{12} r_S^6 \bigl(5912 r^8 - 30952 r^7 r_S + 69997 r^6 r_S^2 - 89072 r^5 r_S^3 + 69736 r^4 r_S^4 \notag \\
&\qquad - 35008 r^3 r_S^5 + 11952 r^2 r_S^6 - 3072 r r_S^7 + 512 r_S^8\bigr) \notag \\
&\quad - 497664 \beta^3 r^{16} r_S^4 \bigl(2768 r^8 - 14816 r^7 r_S + 34858 r^6 r_S^2 - 46975 r^5 r_S^3 + 38912 r^4 r_S^4 \notag \\
&\qquad - 19068 r^3 r_S^5 + 4288 r^2 r_S^6 + 288 r r_S^7 - 256 r_S^8\bigr) \notag \\
&\quad + 2985984 \beta^2 r^{20} r_S^2 \bigl(4032 r^8 - 23872 r^7 r_S + 66312 r^6 r_S^2 - 114896 r^5 r_S^3 + 135697 r^4 r_S^4 \notag \\
&\qquad - 109616 r^3 r_S^5 + 57624 r^2 r_S^6 - 17728 r r_S^7 + 2448 r_S^8\bigr) \notag \\
&\quad + 1146617856 \beta r^{25} r_S^2 (r-r_S)^2 \bigl(42 r^3 - 109 r^2 r_S + 80 r r_S^2 - 12 r_S^3\bigr) \notag \\
&\quad + 330225942528 r^{30} (r-r_S)^4
\Biggr)
\end{align}

    \section{Modified Einstein equations for spherically symmetric space time in  Lemaître–Tolman–Bondi cordinates}   \label{sec:app3}
    By substituting the metric anzats in equation \eqref{eq:metric2} we obtain the modified equations of motion\eqref{eq:modfried} in Lemaitre coordinates. The lapse function $g$ is assumed to be unity. Since the coupling $\beta G$ is small, we expand the equations in $\beta G$ and truncate them at first order. The expanded equations of motion are presented here in this appendix. The equations can be divided in to Einstein tensor components ($G_{\mu\nu}$) without modified terms appearing at 0th order in the expansion and modified equations appearing at first order in coupling. 

\begin{align}
    G_{rr} &= -\frac{3 (\partial_r f)^2 \left[2 f (\partial_{tt} f) + (\partial_t f)^2\right]}{f^2} \\
    H_{rr} &= \frac{1}{8 f^4 \partial_r f} \Bigg\{ 
    -12 f^4 (\partial_r f)^2 (\partial_{r tttt} f) 
    + \Big[24 f^4 (\partial_r f) (\partial_{rt} f) - 48 (\partial_r f)^2 f^3 (\partial_t f)\Big] (\partial_{r ttt} f) \nonumber\\
    &\quad + 4 (\partial_{rrtt} f) f^3 (\partial_r f) - 4 f^3 (\partial_r f)^3 (\partial_{tttt} f) + 17 f^4 (\partial_{rtt} f)^2 \nonumber\\
    &\quad + \Big[ -24 f^4 (\partial_{rt} f)^2 + 66 f^3 (\partial_t f) (\partial_r f) (\partial_{rt} f) - 38 (\partial_r f)^2 f^3 (\partial_{tt} f) \nonumber\\
    &\qquad - 4 f^2 \Big(f (\partial_{rr} f) + (\partial_r f)^2 ((\partial_t f)^2 - 2)\Big) \Big] (\partial_{rtt} f) 
    + \Big[-4 (\partial_r f)^2 f^3 (\partial_{rt} f) - 4 f^2 (\partial_r f)^3 (\partial_t f)\Big] (\partial_{ttt} f) \nonumber\\
    &\quad + 28 f^2 (\partial_r f) (\partial_t f) (\partial_{rrt} f) - 8 f^3 (\partial_t f) (\partial_{rt} f)^3 
    + 8 f^2 (\partial_r f) \left[(\partial_{tt} f) - \frac{5}{8}(\partial_t f)^2 + \frac{7}{2}\right] (\partial_{rt} f)^2 \nonumber\\
    &\quad + \Big[26 (\partial_r f)^2 f^2 (\partial_{tt} f) (\partial_t f) - 10 f (\partial_r f)^2 (\partial_t f)^3 - 28 f^2 (\partial_r f) (\partial_t f) (\partial_{rr} f)\Big] (\partial_{rt} f) \nonumber\\
    &\quad - 45 (\partial_r f)^3 \bigg[ f^2 (\partial_{tt} f)^2 + \frac{2}{5} f (\partial_t f)^2 + \frac{8}{45} f (\partial_{tt} f) + \frac{19}{45} (\partial_t f)^4 + \frac{28}{45} (\partial_t f)^2 \bigg] 
    \Bigg\}
    \label{eq:mo1}
\end{align}

\begin{align}
    G_{\theta \theta}^{(2)} &= -\frac{1}{\partial_r f} \left[ 3 f \left( (\partial_{rtt} f) f + (\partial_t f) (\partial_{rt} f) + (\partial_{tt} f) (\partial_r f) \right) \right] \\
    H_{\theta \theta} &= \frac{1}{8 f^2 (\partial_r f)^5} \Bigg\{
    \Big[ 2 f^4 (\partial_{rrt} f) (\partial_r f)^2 - 2 f^4 (\partial_r f)^4 (\partial_{rt^4} f) 
    + \left(-6 f^4 (\partial_r f) (\partial_{rr} f) + 6 f^3 (\partial_r f)^3 \right) (\partial_{rrtt} f) \Big] \nonumber\\
    &\quad + \Big[ 2 f^4 (\partial_r f)^3 (\partial_{rt} f) - 6 f^3 (\partial_r f)^4 (\partial_t f) \Big] (\partial_{rttt} f) 
    + 14 f^3 (\partial_t f) (\partial_r f)^2 (\partial_{rrt} f) - 14 f^3 (\partial_r f)^5 (\partial_{tttt} f) \nonumber\\
    &\quad - 13 f^4 (\partial_{rtt} f)^2 (\partial_r f)^3 
    - 2 f \bigg[ f^2 (\partial_r f)^2 (\partial_{rt} f)^2 + 19 (\partial_r f)^4 f (\partial_{tt} f) - 5 f (\partial_r f)^3 (\partial_t f) (\partial_{rt} f) \nonumber\\
    &\qquad + (\partial_r f) f^2 (\partial_{rrt} f) - 3 f^2 (\partial_{rr} f)^2 + 3 (\partial_r f)^2 f (\partial_{rr} f) + 2 (\partial_r f)^4 \bigg] (\partial_{rtt} f) \nonumber\\
    &\quad - 14 f^3 (\partial_r f) (\partial_t f) (\partial_{rrt} f) (\partial_{rt} f) 
    + 42 f^3 (\partial_r f) \Big[ (\partial_r f)(\partial_{rt} f) - (\partial_t f)(\partial_{rr} f) \Big] (\partial_{rrt} f) \nonumber\\
    &\quad - 28 f^3 (\partial_r f)^4 (\partial_{ttt} f) (\partial_{rt} f) - 14 f^3 (\partial_r f)^2 (\partial_t f) (\partial_{rt} f)^3 \nonumber\\
    &\quad + \bigg( 14 f^3 (\partial_r f)^3 (\partial_{tt} f) - 15 (\partial_r f)^3 (\partial_t f)^2 f^2 - 42 f^3 (\partial_r f) (\partial_{rr} f) (\partial_{rt} f)^2 
    + 42 (\partial_t f) f^2 (\partial_{rr} f)^2 - \frac{5}{21} (\partial_r f)^4 f (\partial_{tt} f) \nonumber\\
    &\qquad - \left( (\partial_t f)^2 + \frac{14}{5} \right) f (\partial_r f) (\partial_{tt} f) - 15 f^2 (\partial_{rr} f)^2 
    + \left( -\frac{14}{5} f (\partial_t f)^2 - \frac{4}{15} f \right) (\partial_{tt} f) + \frac{23}{15} (\partial_t f)^4 - \frac{28}{15} (\partial_t f)^2 \bigg)
    \Bigg\}
    \label{eq:mo2}
\end{align}

\begin{align}
    G_{t t} &= \frac{3 (\partial_t f) \left[ 2 f (\partial_{rt} f) + (\partial_t f) (\partial_r f) \right]}{f^2 \partial_r f} \\
    H_{t t} &= \frac{1}{8 f^4 (\partial_r f)^5} \Bigg\{
    -12 f^4 (\partial_{rrt} f) (\partial_r f)^2 (\partial_{rtt} f) 
    + \Big( 36 f^4 (\partial_r f) (\partial_{rr} f) - 48 f^3 (\partial_r f)^3 \Big) (\partial_{rrtt} f) \nonumber\\
    &\quad + \Big( 12 f^4 (\partial_r f)^3 (\partial_{rt} f) + 4 f^3 (\partial_r f)^4 (\partial_t f) \Big) (\partial_{rttt} f) 
    - 4 f^3 (\partial_t f) (\partial_r f)^2 (\partial_{rrt} f) - 5 f^4 (\partial_{rtt} f)^2 (\partial_r f)^3 \nonumber\\
    &\quad - 12 f^2 (\partial_r f)^2 (\partial_{rt} f)^2 + \frac{5}{6} (\partial_r f)^4 f (\partial_{tt} f) 
    - \frac{5}{2} f (\partial_r f)^3 (\partial_t f) (\partial_{rt} f) - (\partial_t f)^2 (\partial_r f)^4 \nonumber\\
    &\quad - \Big[ (\partial_r f) f^2 (\partial_{rrt} f) + 3 f^2 (\partial_{rr} f)^2 - 4 (\partial_r f)^2 f (\partial_{rr} f) \Big] (\partial_{rtt} f) \nonumber\\
    &\quad + \Big[ -12 (\partial_r f)^2 f^3 (\partial_{rt} f) + 12 f^3 (\partial_r f) (\partial_t f) (\partial_{rr} f) - 4 f^2 (\partial_r f)^3 (\partial_t f) \Big] (\partial_{rrt} f) \nonumber\\
    &\quad + 4 f^3 (\partial_r f) (\partial_t f) (\partial_{rrt} f) (\partial_{rt} f) + 4 f^3 (\partial_r f)^4 (\partial_{rt} f) (\partial_{t} f) 
    + 28 f^2 (\partial_r f)^5 (\partial_t f) (\partial_{rttt} f) - 4 f^3 (\partial_r f)^2 (\partial_t f) (\partial_{rt} f)^3 \nonumber\\
    &\quad + 4 (\partial_r f) \bigg[ 3 f (\partial_{rr} f) + (\partial_r f)^2 f (\partial_{tt} f) - \left( \frac{7}{4}(\partial_t f)^2 - 1 \right) f^2 (\partial_{rt} f)^2 
    - 12 f^2 (\partial_{rr} f)^2 - \frac{1}{3} (\partial_r f)^2 f (\partial_{rr} f) \nonumber\\
    &\qquad - \left( \frac{23}{6} (\partial_r f)^4 f (\partial_{tt} f) + \left( \frac{9}{23} (\partial_t f)^2 + \frac{4}{23} \right) 6 (\partial_t f) f (\partial_{rt} f) \right) \nonumber\\
    &\qquad - 15 (\partial_r f)^5 f^2 (\partial_{tt} f)^2 - \left( \frac{2}{3} f (\partial_t f)^2 (\partial_{tt} f)^3 \right) 
    + \frac{13}{15} (\partial_t f)^4 + \frac{4}{15} (\partial_t f)^2 \bigg]
    \Bigg\}
    \label{eq:mo3}
\end{align}
\subsection{Reintroducing Lapse function to Einstein equations}
\label{sec:app31}
In general the lapse function is nonzero and can contribute to the Einstein equations at first order in coupling. Contribution of lapse function to the modified equations are of higher order and ignored henceforth. We perform this two stage derivation of equations of motion, since deriving the modified equations correctly, assumes a diagonal Ricci tensor which is not the case if $\frac{\partial g}{\partial r}$ is non zero. Upon reintroducing lapse, the Einstein equations are 

    \begin{equation}
        G_{rr} = -\frac{\frac{\partial f}{\partial r}}{f^2 g^2} \left[ 2f \frac{\partial^2 f}{\partial t^2} \frac{\partial f}{\partial r} g - f \frac{\partial f}{\partial t} \frac{\partial g}{\partial t} \frac{\partial f}{\partial r} + \left( \frac{\partial f}{\partial t} \right)^2 \frac{\partial f}{\partial r} g - f \frac{\partial g}{\partial r} g \right]
    \end{equation}

    \begin{equation}
        G_{\theta \theta} = \frac{3}{4 \left(\frac{\partial f}{\partial r}\right)^3 g^2} \left[ f \left( \begin{aligned} 
        &-4 \frac{\partial^3 f}{\partial r \partial t^2} \left(\frac{\partial f}{\partial r}\right)^2 g f - 4 \frac{\partial^2 f}{\partial t^2} \left(\frac{\partial f}{\partial r}\right)^3 g \\ 
        &+ 2 \frac{\partial g}{\partial t} \frac{\partial f}{\partial t} \left(\frac{\partial f}{\partial r}\right)^3 + 2 \frac{\partial g}{\partial t} \frac{\partial^2 f}{\partial r \partial t} f \left(\frac{\partial f}{\partial r}\right)^2 \\ 
        &- 4 \frac{\partial f}{\partial t} \frac{\partial^2 f}{\partial r \partial t} g \left(\frac{\partial f}{\partial r}\right)^2 + 2 \frac{\partial^2 g}{\partial r^2} \frac{\partial f}{\partial r} g f \\ 
        &- \left(\frac{\partial g}{\partial r}\right)^2 \frac{\partial f}{\partial r} f - 2 \frac{\partial g}{\partial r} \frac{\partial^2 f}{\partial r^2} g f + 2 \frac{\partial g}{\partial r} g \left(\frac{\partial f}{\partial r}\right)^2 
        \end{aligned} \right) \right]
    \end{equation}
    \begin{equation}
        G_{t t} = \frac{3 \frac{\partial f}{\partial t} \left( \frac{\partial f}{\partial t} \frac{\partial f}{\partial r} + 2 f \frac{\partial^2 f}{\partial r \partial t} \right)}{\frac{\partial f}{\partial r} f^2}
    \end{equation}
\end{document}